\documentclass[aps,pra,twocolumn,showpacs]{revtex4}
\usepackage{amsmath,amssymb,graphicx,bbold}
\usepackage{dcolumn}   
\usepackage{bm}        
\usepackage{color}

\begin{document}

\title{Spin-orbit laser mode transfer via a classical analogue of quantum teleportation}
\author{B. Pinheiro da Silva, M. Astigarreta Leal, C. E. R. Souza, E. F. Galv\~ao and A. Z. Khoury}

\affiliation{
Instituto de F\'\i sica, Universidade Federal Fluminense,
24210-346 Niter\'oi - RJ, Brasil}
\date{\today}

\begin{abstract}
We translate the quantum teleportation protocol into a sequence of coherent operations 
involving three degrees of freedom of a classical laser beam. The protocol, which we 
demonstrate experimentally, transfers the polarisation state of the input beam to the 
transverse mode of the output beam. The role of quantum entanglement is played by a 
non-separable mode describing the path and transverse degrees of freedom. Our protocol 
illustrates the possibility of new optical applications based on this intriguing 
classical analogue of quantum entanglement.
\end{abstract}

\maketitle

\section{introduction}
\label{intro}

The tensor product structure of the vector space describing composite quantum systems 
is the key to the very definition of quantum entanglement. It was noticed many years ago 
\cite{spreeuw} that the same mathematical structure arises in the description of classical 
optical fields, yielding what has been called classical entanglement or non-separability. 
This analogue of quantum entanglement allows us to translate quantum information concepts 
to the classical domain. Our first contribution to this topic was the demonstration of a 
topological phase acquired by maximally entangled qubits in the spin-orbit modes of a 
classical laser beam \cite{milman, topoluff}. Later, we investigated a Bell inequality to 
characterize the nonseparability between polarization and the spatial mode structure of 
classical paraxial beams \cite{bell2}. This investigation was also later taken up in the 
single photon regime by other groups \cite{chen, bellpadgett}.
Recently, the role of Bell inequalities in classical optics has drawn a fair amount of 
attention, being addressed in a series of papers \cite{kagalwala, eberly, eberly2}. 
This new understanding of classical non-separability resulted in new optical applications 
inspired by quantum information 
\cite{simon, aiello1, aiello2, ghose, aiello3, forbes, forbes2, shg}. 

The possibility of using the polarization degree of freedom to control spatial modes has 
resulted in a number of applications both in classical and quantum optics. Examples include 
quantum image control \cite{imagepol}, spin to orbital degree of freedom information transfer 
\cite{quplate, chen2009, barreiro2010}, quantum cryptography 
\cite{cryptosteve, cryptouff, cryptolorenzo}, controlled gates \cite{cnotsteve, cnotuff}, 
quantum games \cite{josabuff}, environment-induced entanglement \cite{environ}, quantum 
teleportation \cite{khourymilman}. The coherent superposition of different transverse modes 
carrying orthogonal polarizations creates polarization vortices that have been proved useful 
for classical and quantum encoding of information \cite{cardano, milione, milione2}. 

One of the chief uses of quantum entanglement is information distribution based on quantum 
teleportation \cite{TELE}. It enables one party (Alice), to transfer an arbitrary, unknown 
quantum state to a second party (Bob) via the use of previously shared quantum entanglement 
and classical communication \cite{DEMARTINI, telephoton, teleion, teleQED, telereview}. 
An investigation of an optical analogue of the teleportation protocol has been recently 
reported in \cite{rafsanjani}, in which states of orbital angular momentum were transferred 
to polarization. 

Here we report on experiments in which we perform the interferometric transfer of arbitrary 
polarization modes to the transverse spatial structure of a paraxial, classical laser beam. 
This is achieved  by mapping the steps of a single-qubit quantum teleportation protocol into 
a sequence of controlled optical mode operations. In our experiments, the role of quantum 
entanglement is played by a classical non-separable joint state of the path and transverse 
spatial degrees of freedom of the laser beam. Our protocol translates the relative ease of 
polarization preparation into the possibility of high-quality preparation of arbitrary states 
of transverse modes of a laser beam. Our results push further the analogy between classical 
and quantum entanglement, showing new optical applications are possible if we explore this 
analogy in depth. 

Our paper is organized as follows. In section \ref{formal} we describe the optical modes we 
use to describe the different degrees-of-freedom we manipulate in our experiment. 
In section \ref{protocol} we translate the steps of the quantum teleportation protocol to a 
series of operations on a classical laser beam. In section \ref{experiment} we describe the 
experimental setup and results, with some concluding remarks in section \ref{conclusion}. 

\section{Optical mode structure}
\label{formal}

Our experiment involves interference and polarization measurements on first-order paraxial 
beams. We will work in the computational basis of Laguerre-Gaussian modes of width $w\,$, 
carrying orbital angular momentum. 
Inside an interferometer, these modes are distributed between two 
alternative paths determined by the longitudinal propagation axes of the modes. 
For simplicity, we assume two axes lying parallel to the $z$ direction of the coordinate
system, lying a distance $d\gg w$ apart from each other, so that the overlap between transverse 
modes belonging to different paths is negligible. We shall refer to these paths as $0$ and $1$ 
and designate their locations on the $x-y$ plane by the coordinates $(0,0)$ and $(d,0)\,$, 
respectively.

\begin{figure}[h!]
\includegraphics[scale=0.5]{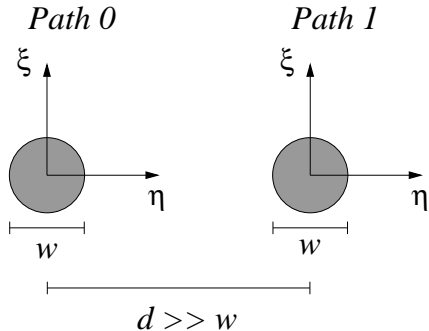}
\caption{\label{modes} Cross section view of the two-path propagation with 
the relative transverse coordinates.}
\end{figure}

We describe the transverse modes in the normalized Laguerre-Gaussian basis and focus on the 
first order, two-dimensional vector space. Their functional form is 
\begin{eqnarray}
\psi_{\pm}(\xi,\eta) = \frac{2}{\sqrt{\pi}}\, \left(\xi \pm i \eta\right)\,
e^{-\left[(\xi^2+\eta^2)(1+i\tilde{z})-i\phi(\tilde{z})\right]}\,,
\end{eqnarray}
where
\begin{eqnarray}
\tilde{z} &=& z/z_0\;,
\nonumber\\
(\xi,\eta) &=& (x,y)/w(\tilde{z}) \;,
\nonumber\\
w(\tilde{z}) &=& w_0 \sqrt{1+\tilde{z}^2}\;,
\nonumber\\
\phi(\tilde{z}) &=& \arctan(\tilde{z})\;,
\end{eqnarray}
$z_0 = k\,w_0^2/2$ is the Rayleigh distance and $w_0$ is the beam waist. The transverse 
coordinates $(\xi,\eta)$ are relative to the path axes and normalized by the beam width. 
This arrangement is depicted in Fig.(\ref{modes}).

Transverse modes lying on different paths do not overlap and are orthogonal, regardless 
of their functional form. Therefore, we can formally represent the path modes as an 
independent degree of freedom and describe them by a pair of column vectors 
\begin{eqnarray}
\chi_{0} &=& \left[
\begin{array}{l}
1\\
0 
\end{array}\right]\,,\quad
\chi_{1} = \left[
\begin{array}{l}
0\\
1 
\end{array}\right]
\;. 
\label{colunas}
\end{eqnarray}
We then build the path-transverse mode structure as a tensor product 
$\{\psi_+,\psi_-\}\otimes\{\chi_0,\chi_1\}\,$. Finally, each transverse 
mode lying on each path has two possible polarizations described by 
horizontal and vertical unit vectors $\{\hat{e}_H,\hat{e}_V\}\,$. Thus, 
our complete workspace will be 
\begin{eqnarray}
\mathcal{W} = \{\psi_+,\psi_-\}\otimes\{\chi_0,\chi_1\}\otimes\{\hat{e}_H,\hat{e}_V\}\,. 
\end{eqnarray}  

These three degrees of freedom can be operated separately or in a combined way in 
order to implement controlled operations. In the polarization degree of freedom, 
unitary transformations are implemented with waveplates, and projections onto linear basis 
vectors are performed by polarizers. Transverse modes can be transformed by astigmatic mode 
converters or Dove prisms. Path modes can be operated on with beam splitters and 
phase shifters. Controlled operations are easily performed when transverse 
mode or polarization transformations are applied on one path only. With this set of 
operations we will implement the steps of a quantum teleportation protocol to 
transfer an arbitrary polarization superposition to the transverse mode, using the resource 
of classical, non-separable states of the path and transverse degrees of freedom.


\section{The protocol}
\label{protocol}

The teleportation protocol involves three steps. First, an arbitrary polarization mode 
is prepared on the laser beam impinging on path $0\,$. Then, the transverse mode is 
entangled with the path degree of freedom with a conditional operation. Finally, a Bell 
projection is performed on the path and polarization degrees of freedom, resulting in four 
transverse mode superpositions at the different outputs. One of them 
(the \textit{successful} output) will carry a transversal mode state which directly 
corresponds to the initial polarisation state. The other three outputs will correspond 
to the initial polarisation state changed by Pauli operations, in direct correspondence 
with the one-qubit quantum teleportation protocol. In what follows we describe each step 
in more detail.

\subsection{Polarization preparation}

\begin{figure}[h!]
\includegraphics[scale=0.4]{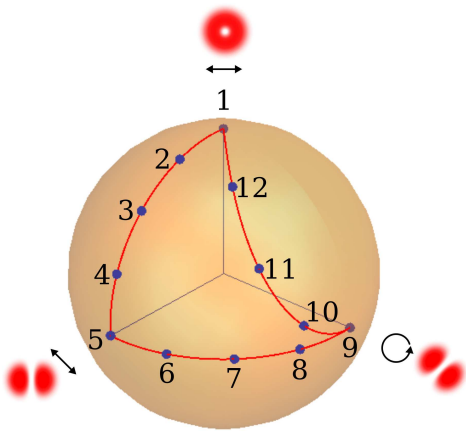}
\caption{\label{sphere} Bloch sphere representation of the input polarization modes 
used in the teleportation protocol together with the corresponding output transverse modes.}
\end{figure}

In order to demonstrate the protocol, we sent a sequence of twelve different polarization 
modes prepared with a half and a quarter waveplates. This sequence is represented in the 
Bloch sphere shown in Fig.(\ref{sphere}), where the corresponding transverse modes 
produced are also represented. 
By rotating the orientations of the waveplates, we prepare polarisation states which 
interpolate between three mutually unbiased 
modes: \textit{i-}) linear horizontal (point $1$), \textit{ii-}) linear at $45^\circ$ (point $5$) 
and \textit{iii-}) left circular (point $9$). Each point corresponds to a polarization suprposition  
\begin{eqnarray}
\hat{\varphi}_n = \alpha_n\,\hat{e}_H + \beta_n\,\hat{e}_V\;\;\; (1\leq n\leq 12)\;,
\end{eqnarray}
with $\alpha_n$ and $\beta_n$ determined by the polar and azimuthal angles on the sphere. 

\subsection{Path-transverse mode entanglement}

\begin{figure}[h!]
\includegraphics[scale=0.4]{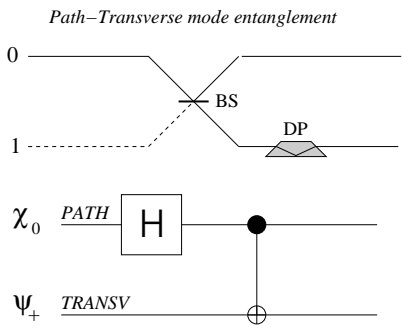}
\caption{\label{esquema-preparation} Top: Path-transverse mode entanglement scheme. 
BS - beam splitter, DP - Dove prism. Bottom: Equivalent quantum circuit representation.}
\end{figure}

A non-separable path-transverse mode state can be created by applying a path Hadamard 
operation followed by a conditional transverse mode flip. This is 
achieved with a Dove prism inserted on path $1$ after the beam splitter, as sketched in 
Fig.(\ref{esquema-preparation}). 

\subsection{Path-polarization Bell projection}

\begin{figure}[h!]
\includegraphics[scale=0.4]{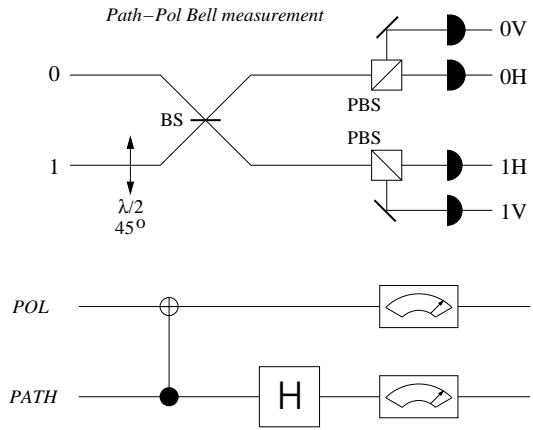}
\caption{\label{esquema-Bell} Top: Path-polarization Bell projection scheme. Bottom: Equivalent 
quantum circuit representation.}
\end{figure}

An essential ingredient of the protocol is the ability to perform a Bell measurement in the 
degrees of freedom corresponding to the polarisation and beam path. This can be done via a 
controlled unitary gate between them, followed by a projection on single-system bases. 
In our scheme, we make a path dependent polarization transformation followed by a beam 
splitter operation (which corresponds to a Hadamard gate on the path mode), followed by 
polarization measurements with polarizing beam splitters. 
The path-polarization Bell measurement scheme is described at the top of 
Fig.(\ref{esquema-Bell}). The controlled polarization transformation is implemented with a 
half waveplate oriented at $45^\circ\,$, placed on path $1\,$. This corresponds to a CNOT gate 
that flips the polarization (target) conditioned to the path mode (control). 
Then, the two path modes are combined in a $50/50$ beam splitter, corresponding to a 
Hadamard gate. One can easily verify that this sequence transforms the four 
path-polarization Bell modes as follows:
\begin{eqnarray}
\left(\chi_0\,\hat{e}_H + \chi_1\,\hat{e}_V\right)/\sqrt{2} \rightarrow \chi_0\,\hat{e}_H\;,
\nonumber\\
\left(\chi_0\,\hat{e}_H - \chi_1\,\hat{e}_V\right)/\sqrt{2} \rightarrow \chi_1\,\hat{e}_H\;,
\nonumber\\
\left(\chi_0\,\hat{e}_V + \chi_1\,\hat{e}_H\right)/\sqrt{2} \rightarrow \chi_0\,\hat{e}_V\;,
\nonumber\\
\left(\chi_0\,\hat{e}_V - \chi_1\,\hat{e}_H\right)/\sqrt{2} \rightarrow \chi_1\,\hat{e}_V\;. 
\label{belltransform}
\end{eqnarray}
Finally, a polarization measurement performed at each output path can discriminate the 
four input Bell modes. These operations can be summarized as a quantum circuit, as shown 
at the bottom of Fig.(\ref{esquema-Bell}). 

\section{Experimental setup and results}
\label{experiment}

\begin{figure}[h!]
\includegraphics[scale=0.33]{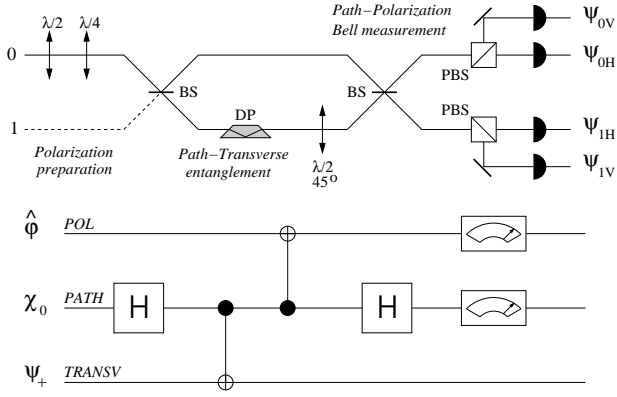}
\caption{\label{setup} Top: Experimental scheme for spin-orbit mode teleportation. 
Bottom: Equivalent quantum circuit representation.}
\end{figure}

The experimental setup is described in Fig.(\ref{setup}). A horizontally polarized $TEM_{00}$ mode 
produced by a He-Ne laser is diffracted by a holographic mask to prepare a 
Laguerre-Gaussian transverse mode $\psi_+\,$ propagating on path $\chi_0\,$. 
Two waveplates convert its polarization to a chosen superposition 
\begin{eqnarray}
\hat{\varphi} = \alpha\,\hat{e}_H + \beta\,\hat{e}_V\;,
\end{eqnarray}
thus preparing the initial separable supermode 
\begin{eqnarray}
\mathbf{\Psi}_A = \psi_+ (\eta,\xi)\,\chi_0\,\hat{\varphi}\;.
\end{eqnarray}
For simplicity, we shall omit the arguments in the transverse modes from now on. 
A beam splitter (BS) is used to perform a Hadamard gate in the path mode, and 
a Dove prism is inserted in output path $1$ of the BS to flip the transverse mode 
propagating on this path. This operation is modelled by a CNOT gate on the transverse 
mode (target), controlled by the path; it creates the non-separable transverse-mode/path 
state which plays the role of a pair of maximally entangled qubits in the quantum 
teleportation protocol. This supermode is described by:
%
\begin{eqnarray}
\mathbf{\Psi}_B = \left[\frac{\psi_+ \,\chi_0  + 
\psi_- \,\chi_1 }{\sqrt{2}}\right] \,\hat{\varphi}\;.
\end{eqnarray}

After the controlled operation between path and transverse mode, we start the sequence 
of operations that will perform the Bell projection on the path and
polarization modes, as described in section \ref{formal}. First, a second 
controlled gate is performed by a half-wave plate oriented at $45^o$ (Pauli $\sigma_X$ 
operation on polarization), inserted on path $1\,$. It is equivalent to a CNOT gate on 
the polarization mode (target) controlled by the path, producing the supermode
\begin{eqnarray}
\mathbf{\Psi}_C &=& 
\frac{\psi_+ \,\chi_0\,\left(\alpha\,\hat{e}_H + \beta\,\hat{e}_V\right)  + 
\psi_- \,\chi_1\,\left(\beta\,\hat{e}_H + \alpha\,\hat{e}_V\right) }{\sqrt{2}}\;.
\nonumber\\ 
\end{eqnarray}
Then, a Hadamard operation is performed on the path degree-of-freedom by a beam splitter, 
resulting in the output supermode
\begin{eqnarray}
\mathbf{\Psi}_D &=& 
\left[\frac{}{}\left(\alpha\,\psi_+  + \beta\,\psi_-\right)\,\chi_0\,\hat{e}_H\right.
\nonumber\\
&+&
\left(\beta\,\psi_+  + \alpha\,\psi_-\right)\,\chi_0\,\hat{e}_V
\nonumber\\
&+&
\left(\alpha\,\psi_+  - \beta\,\psi_-\right)\,\chi_1\,\hat{e}_H
\nonumber\\
&+&
\left.\left(\beta\,\psi_+  - \alpha\,\psi_-\right)\,\chi_1\,\hat{e}_V\frac{}{}\right]/2
\;.
\end{eqnarray}
To complete the Bell measurement a polarization projection is performed with a polarizing 
beam splitter placed in each output path, giving output transverse modes
\begin{eqnarray}
\psi_{ij} (\eta,\xi) &=& 
\left(\chi_i^\dagger\,\hat{e}^*_j\right)\cdot \mathbf{\Psi}_D\;,
\end{eqnarray}
with $i=0,1$ and $j=H,V\,$. The four output transverse modes are
\begin{eqnarray}
\psi_{0H} &=& 
\alpha\,\psi_+  + \beta\,\psi_-\;,
\nonumber\\
\psi_{0V} &=& 
\beta\,\psi_+  + \alpha\,\psi_-\;,
\nonumber\\
\psi_{1H} &=& 
\alpha\,\psi_+  - \beta\,\psi_-\;,
\nonumber\\
\psi_{1V} &=& 
\beta\,\psi_+  - \alpha\,\psi_- \;.
\label{outputm}
\end{eqnarray}
Output port $0H$ is the successful one, which does not require any unitary correction, 
while ports $0V\,$, $1H$ and $1V\,$, must be corrected with $\sigma_X\,$, $\sigma_Z$ 
and $\sigma_X\sigma_Z\,$ operations, respectively. 

\begin{figure}[h!]
\includegraphics[scale=0.4]{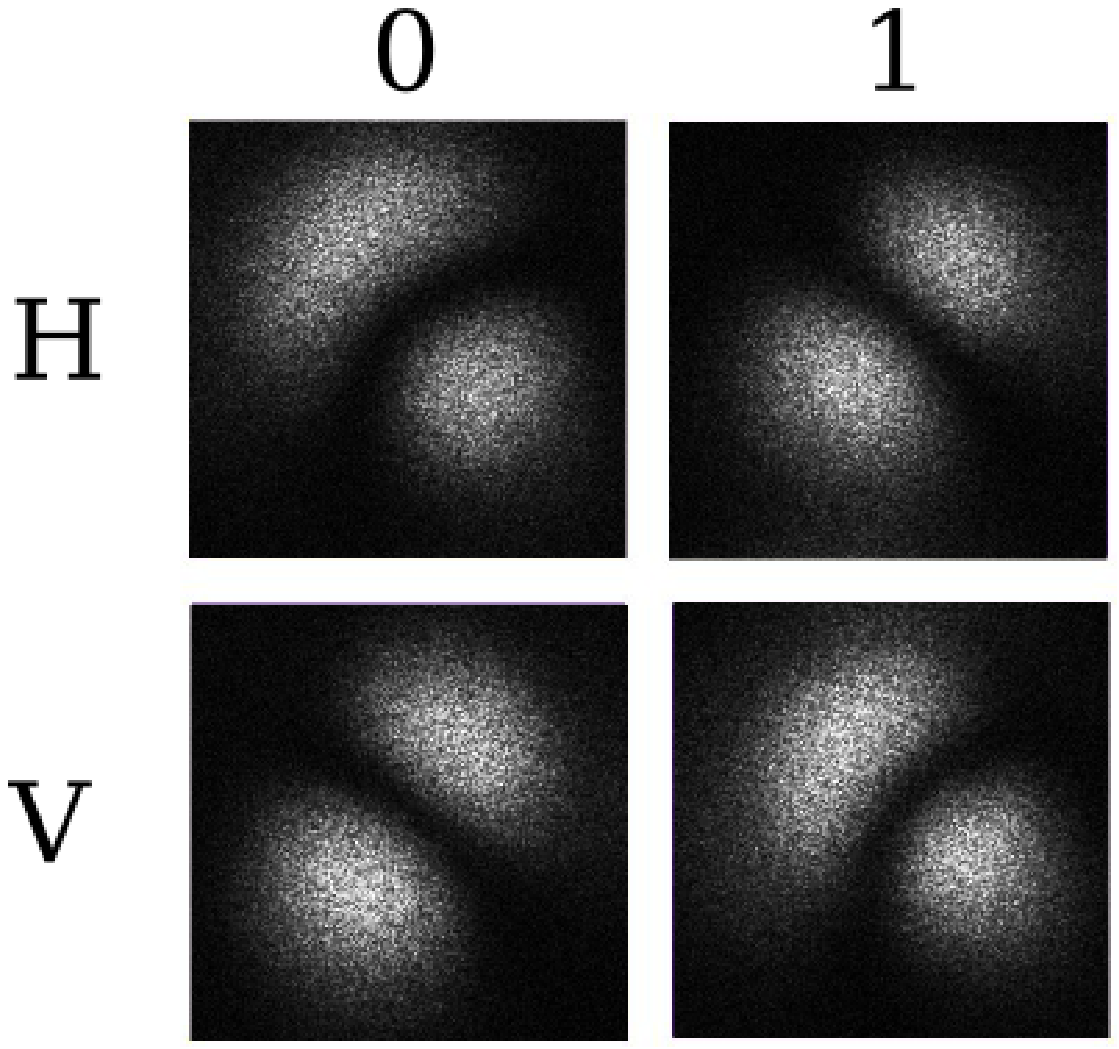}
\includegraphics[scale=0.4]{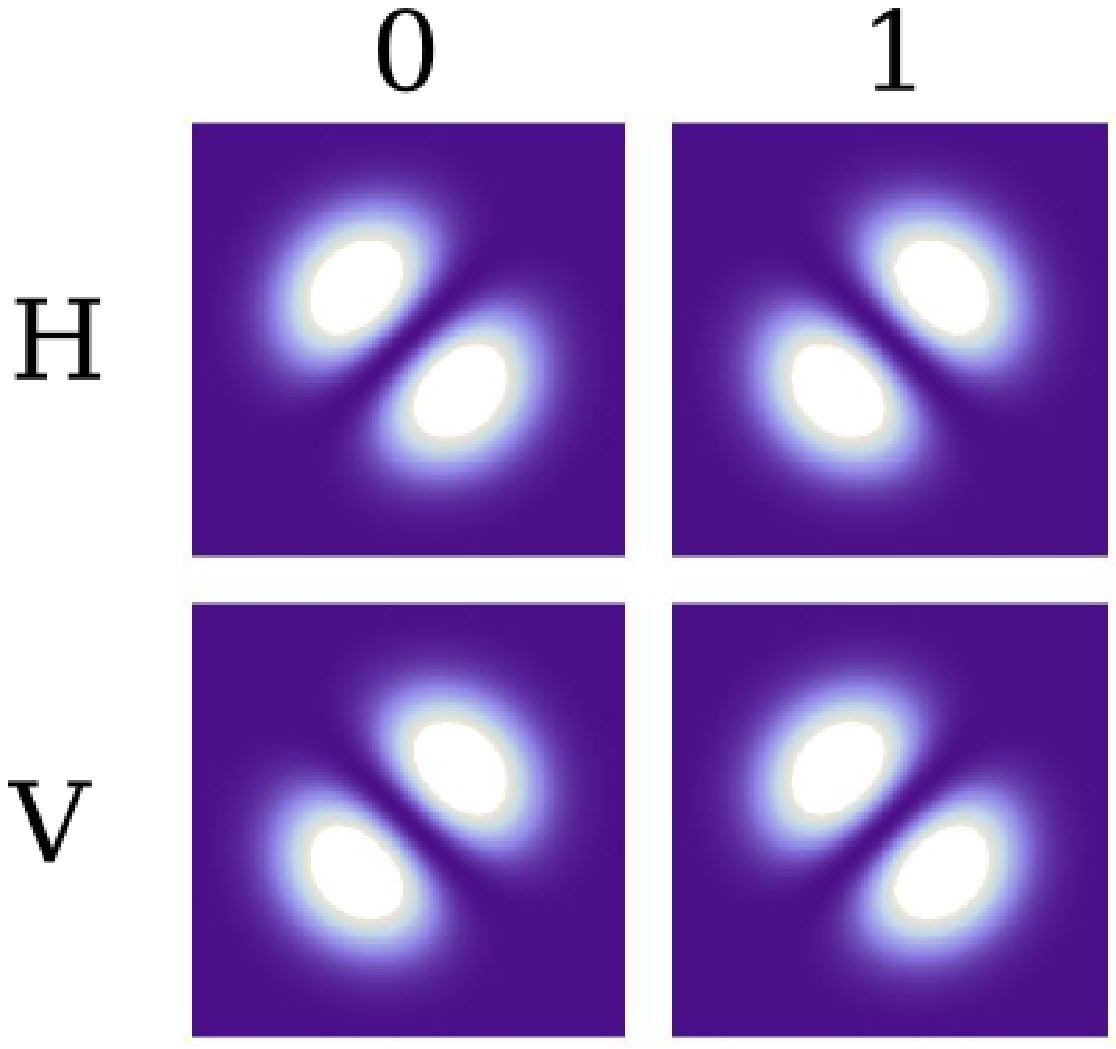}
\caption{\label{4images} Images of the four teleportation outputs for 
a circularly polarized input. Top: Experimental results. Bottom: Numerical simulations.}
\end{figure}

The four outputs were registered with a CCD camera. First we prepared the input beam with
left circularly polarized light, which corresponds to $\alpha = 1/\sqrt{2}\,$ and 
$\beta = i/\sqrt{2}\,$. 
This setting is expected to produce four Hermite-Gaussian beams oriented at 
$-45^{0}$ on port $0H\,$, $45^{0}$ on port $0V\,$, $45^{0}$ on port $1H\,$ and 
$-45^{0}$ on port $1V\,$, which is in very good agreement with the results shown in 
the top four images of Fig.(\ref{4images}). The bottom images are 
theoretical density plots of the transverse mode superpositions given by 
Eq.(\ref{outputm}). 

\begin{figure}[h!]
\includegraphics[scale=0.4]{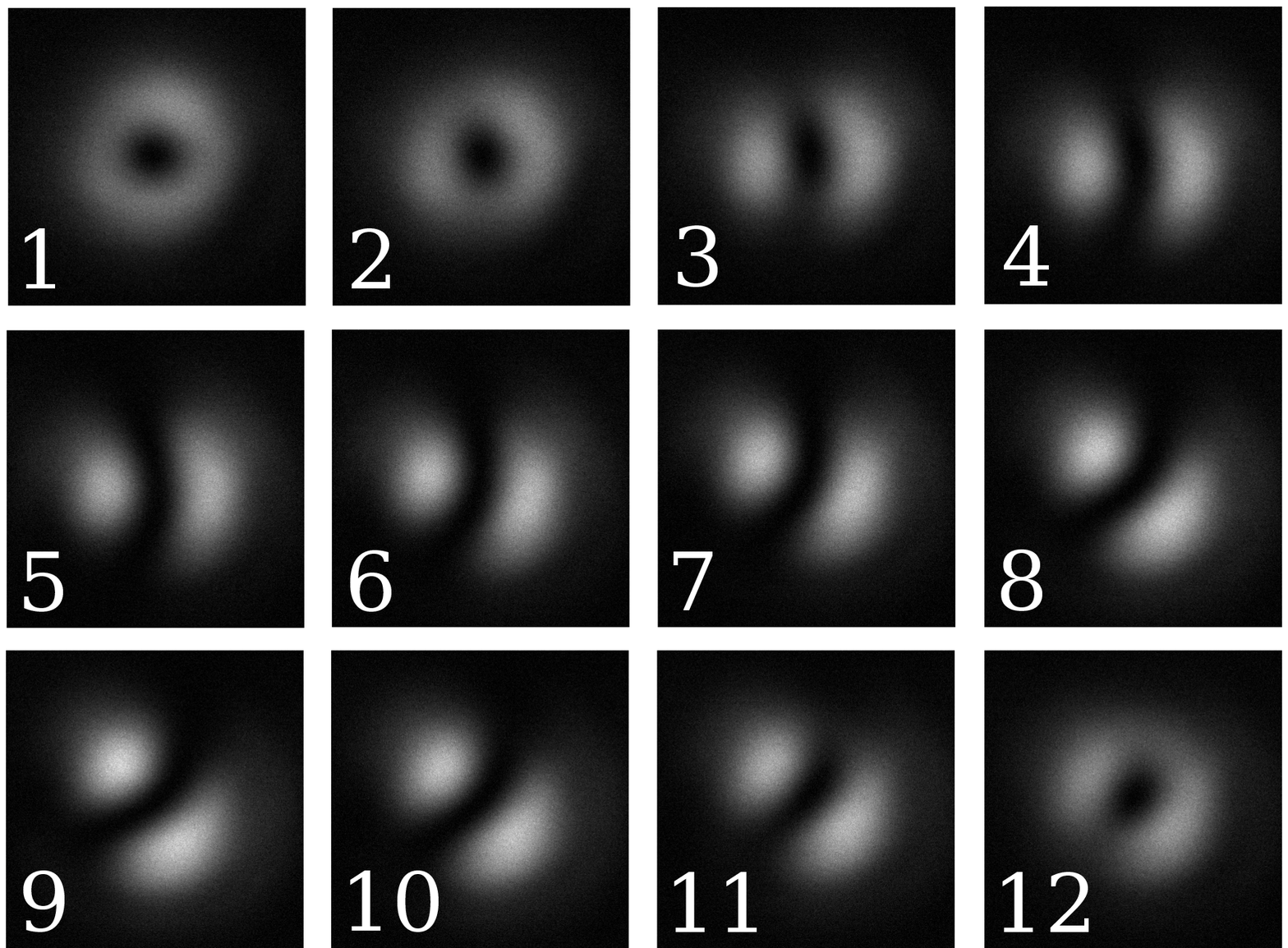}
\includegraphics[scale=0.4]{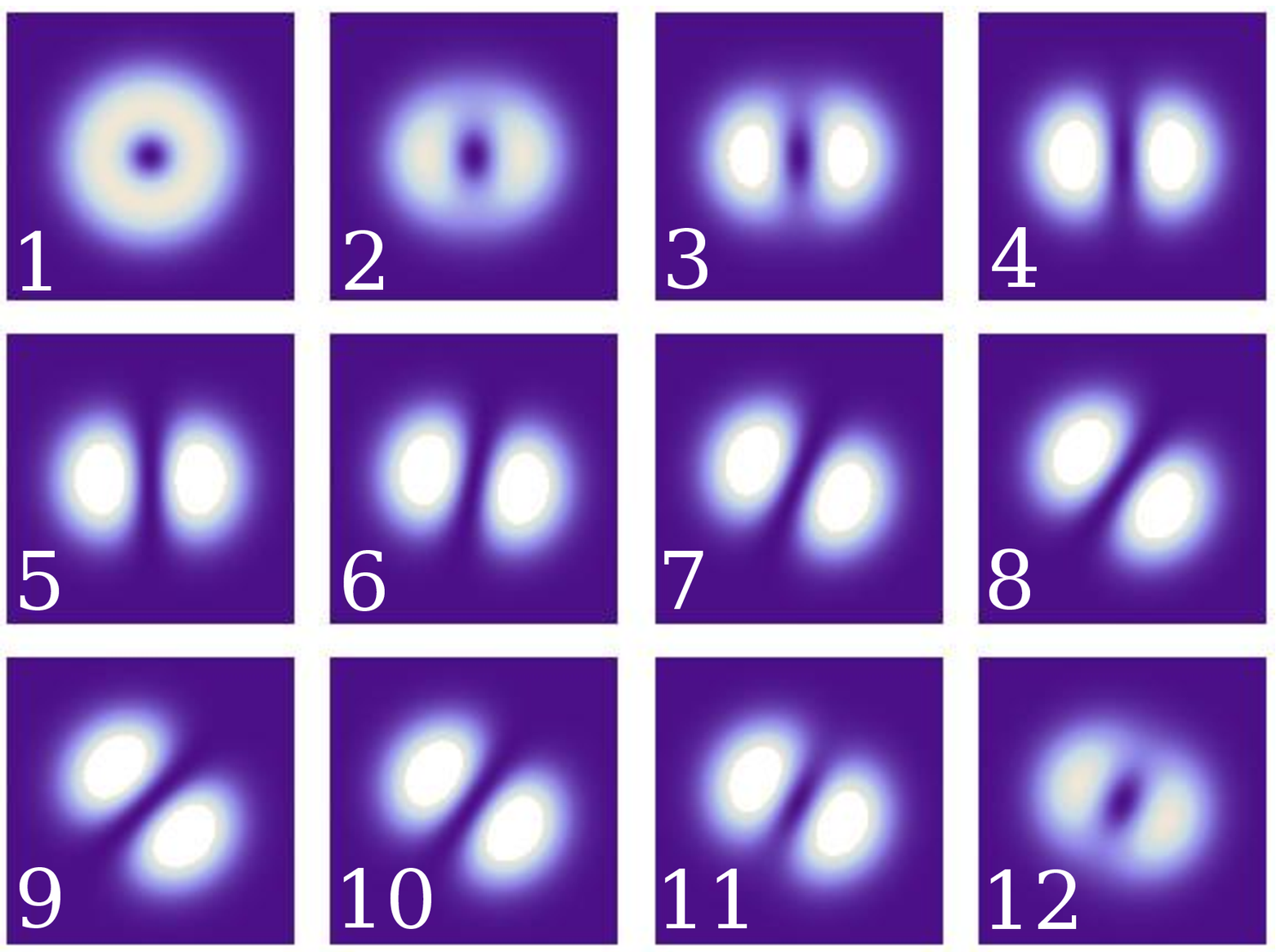}
\caption{\label{poincare-images} Poincar\'e representation of the input polarization modes 
used in the teleportation protocol. Top: Experimental results. Bottom: Numerical simulations.}
\end{figure}

Then, a sequence of polarization modes was prepared by rotating the quarter and 
half waveplates at different orientations. This produced twelve polarization modes, 
forming a closed path in the Poincar\'e sphere, represented in Fig.(\ref{sphere}). 
The images obtained in the successful port for these twelve input polarizations were 
registered on the CCD camera and shown in Fig.(\ref{poincare-images}) together with 
the corresponding theoretical density plots of the expected transverse modes on this 
port. The experimental images are in good agreement with the numerical simulations. 
%


\section{Conclusion}
\label{conclusion}

We have proposed and experimentally demonstrated a scheme to transfer an arbitrary 
polarization state to the first order transverse structure of a paraxial laser beam. 
The scheme mirrors the quantum teleportation protocol and uses operations on three 
internal degrees of freedom of a paraxial laser beam. It can be used as a practical 
way of generating arbitrary first order transverse modes either for classical optical 
processing or quantum cryptography in the photocount regime. A demonstration of a 
protocol similar to the one implemented by us has been independently reported 
in Ref.\cite{szameit}.

\section*{Acknowledgments}
Funding was provided by 
Conselho Nacional de Desenvolvimento Tecnol\'ogico (CNPq), 
Coordena\c c\~{a}o de Aperfei\c coamento de 
Pessoal de N\'\i vel Superior (CAPES), Funda\c c\~{a}o de Amparo \`{a} 
Pesquisa do Estado do Rio de Janeiro (FAPERJ), and Instituto Nacional 
de Ci\^encia e Tecnologia de Informa\c c\~ao Qu\^antica (INCT-CNPq).

\end{document}